\documentclass[proceedings]{stacs}
\stacsheading{2010}{335-346}{Nancy, France}
\firstpageno{335}

\usepackage{verbatim}
\usepackage{amsfonts}
\usepackage{amssymb}
\usepackage{amsmath}
\usepackage{pstricks,pst-node,pst-tree}

\usepackage{graphicx}
\begin{document}

\title{The complexity of the
list homomorphism problem for graphs}

\author[lab1]{L. Egri}{L\'aszl\'o Egri}
\address[lab1]{School of Computer Science, McGill University, Montr\'eal,
Canada}
\email{laszlo.egri@mail.mcgill.ca}

\author[lab2]{A. Krokhin}{Andrei Krokhin}
\address[lab2]{School of Engineering and Computing Sciences, Durham
University, Durham, UK}
\email{andrei.krokhin@durham.ac.uk}

\author[lab3]{B. Larose}{Benoit Larose}
\address[lab3]{Department of Mathematics and Statistics, Concordia
University, Montr\'eal, Canada}
\email{larose@mathstat.concordia.ca}

\author[lab4]{P. Tesson}{Pascal Tesson}
\address[lab4]{Department of Computer Science, Laval University, Quebec City,
Canada}
\email{pascal.tesson@ift.ulaval.ca}


\date{}

\newcommand{\csp}{\ensuremath{\operatorname{CSP}}}
\newcommand{\lcsp}{\ensuremath{\operatorname{LCSP}}}
\newcommand{\LHom}{\ensuremath{\operatorname{LHom}}}
\newcommand{\Hom}{\ensuremath{\operatorname{Hom}}}
\newcommand{\CSP}{\ensuremath{\operatorname{CSP}}}
\newcommand{\xbar}{\bar{x}}
\newcommand{\ybar}{\bar{y}}
\newcommand{\zbar}{\bar{z}}
\newcommand{\wbar}{\bar{w}}
\newcommand{\vbar}{\bar{v}}
\newcommand{\pbar}{\bar{p}}
\newcommand{\qbar}{\bar{q}}
\newcommand{\ubar}{\bar{u}}
\newcommand{\nin}{\ensuremath{\notin}}
\def\zd{,\ldots,}



\newcommand{\cA}{\mathcal{A}}
\newcommand{\cB}{\mathcal{B}}
\newcommand{\ccC}{\mathcal{C}}
\newcommand{\cD}{\mathcal{D}}
\newcommand{\cE}{\mathcal{E}}
\newcommand{\cF}{\mathcal{F}}
\newcommand{\cG}{\mathcal{G}}
\newcommand{\cH}{\mathcal{H}}
\newcommand{\cI}{\mathcal{I}}
\newcommand{\cJ}{\mathcal{J}}
\newcommand{\cK}{\mathcal{K}}
\newcommand{\cL}{\mathcal{L}}
\newcommand{\cM}{\mathcal{M}}
\newcommand{\cN}{\mathcal{N}}
\newcommand{\cO}{\mathcal{O}}
\newcommand{\cP}{\mathcal{P}}
\newcommand{\cQ}{\mathcal{Q}}
\newcommand{\cR}{\mathcal{R}}
\newcommand{\cS}{\mathcal{S}}
\newcommand{\cT}{\mathcal{T}}
\newcommand{\cU}{\mathcal{U}}
\newcommand{\cV}{\mathcal{V}}
\newcommand{\cW}{\mathcal{W}}
\newcommand{\cX}{\mathcal{X}}
\newcommand{\cY}{\mathcal{Y}}
\newcommand{\cZ}{\mathcal{Z}}


\newcommand{\bbA}{\mathbb{A}}
\newcommand{\bbB}{\mathbb{B}}
\newcommand{\bbC}{\mathbb{C}}
\newcommand{\bbD}{\mathbb{D}}
\newcommand{\bbE}{\mathbb{E}}
\newcommand{\bbF}{\mathbb{F}}
\newcommand{\bbG}{\mathbb{G}}
\newcommand{\bbH}{\mathbb{H}}
\newcommand{\bbI}{\mathbb{I}}
\newcommand{\bbJ}{\mathbb{J}}
\newcommand{\bbK}{\mathbb{K}}
\newcommand{\bbL}{\mathbb{L}}
\newcommand{\bbM}{\mathbb{M}}
\newcommand{\bbN}{\mathbb{N}}
\newcommand{\bbO}{\mathbb{O}}
\newcommand{\bbP}{\mathbb{P}}
\newcommand{\bbQ}{\mathbb{Q}}
\newcommand{\bbR}{\mathbb{R}}
\newcommand{\bbS}{\mathbb{S}}
\newcommand{\bbT}{\mathbb{T}}
\newcommand{\bbU}{\mathbb{U}}
\newcommand{\bbV}{\mathbb{V}}
\newcommand{\bbW}{\mathbb{W}}
\newcommand{\bbX}{\mathbb{X}}
\newcommand{\bbY}{\mathbb{Y}}
\newcommand{\bbZ}{\mathbb{Z}}


\newcommand{\bA}{\mathbf{A}}
\newcommand{\bB}{\mathbf{B}}
\newcommand{\bC}{\mathbf{C}}
\newcommand{\bD}{\mathbf{D}}
\newcommand{\bE}{\mathbf{E}}
\newcommand{\bF}{\mathbf{F}}
\newcommand{\bG}{\mathbf{G}}
\newcommand{\bH}{\mathbf{H}}
\newcommand{\bI}{\mathbf{I}}
\newcommand{\bJ}{\mathbf{J}}
\newcommand{\bK}{\mathbf{K}}
\newcommand{\bL}{\mathbf{L}}
\newcommand{\bM}{\mathbf{M}}
\newcommand{\bN}{\mathbf{N}}
\newcommand{\bO}{\mathbf{O}}
\newcommand{\bP}{\mathbf{P}}
\newcommand{\bQ}{\mathbf{Q}}
\newcommand{\bR}{\mathbf{R}}
\newcommand{\bS}{\mathbf{S}}
\newcommand{\bT}{\mathbf{T}}
\newcommand{\bU}{\mathbf{U}}
\newcommand{\bV}{\mathbf{V}}
\newcommand{\bW}{\mathbf{W}}
\newcommand{\bX}{\mathbf{X}}
\newcommand{\bY}{\mathbf{Y}}
\newcommand{\bZ}{\mathbf{Z}}

\newcommand{\datprog}{\ensuremath{\mathcal{D}}}

\newenvironment{proofsketch}{\begin{trivlist} \item {{\em Proof} [sketch]:}}{\hfill $\blacksquare$ \end{trivlist}\vspace*{.1cm}}

\newcounter{k}
\newenvironment{redolemma}[1]{
\setcounter{k}{\value{theorem}}\setcounter{theorem}{#1}
\addtocounter{theorem}{-1}
\begin{lemma}}{\end{lemma} \setcounter{theorem}{\value{k}}}

\newenvironment{redotheorem}[1]{
\setcounter{k}{\value{theorem}}\setcounter{theorem}{#1}
\addtocounter{theorem}{-1}
\begin{theorem}}{\end{theorem} \setcounter{theorem}{\value{k}}}





\begin{abstract}
  \noindent We completely classify the computational complexity of the
  list $\bH$-colouring
problem for graphs (with possible loops) in combinatorial and
algebraic terms: for every graph $\bH$ the problem is either
NP-complete, NL-complete, L-complete or is first-order definable;
descriptive complexity equivalents are given as well via Datalog
and its fragments. Our algebraic characterisations match important
conjectures in the study of constraint satisfaction problems.

\end{abstract}

\maketitle

\section{Introduction}\label{intro}

{\em Homomorphisms of graphs}, i.e.\ edge-preserving mappings, generalise graph
colourings, and can model a wide variety of combinatorial problems dealing with
mappings and assignments~\cite{Hell04:book}. Because of the richness of the
homomorphism framework, many computational aspects of graph homomorphisms have
recently become the focus of much attention. In the {\em list} $\bH$-{\em colouring}
problem (for a fixed graph $\bH$), one is given a graph $\bG$ and a list $L_v$ of
vertices of $\bH$ for each vertex $v$ in $\bG$, and the goal is to determine whether
there is a homomorphism $h$ from $\bG$ to $\bH$ such that $h(v) \in L_v$ for all
$v$. The complexity of such problems has been studied by combinatorial methods,
e.g., in~\cite{Feder99:list,Feder03:bi-arc}. In this paper, we study the complexity
of the list homomorphism problem for graphs in the wider context of classifying the
complexity of constraint satisfaction problems (CSP),
see~\cite{Barto09:digraphs,Feder98:monotone,Hell08:survey}. It is well known that
the CSP can be viewed as the problem of deciding whether there exists a homomorphism
from a relational structure to another, thus naturally extending the graph
homomorphism problem.

One line of CSP research studies the non-uniform CSP, in which the
target (or template) structure ${\bf T}$ is fixed and the question
is whether there exists a homomorphism from an input structure to
$\bf T$. Over the last years, much work has been done on
classifying the complexity of this problem, denoted $\Hom(\bf T)$
or $\csp({\bf T})$, with respect to the fixed target structure,
see
surveys~\cite{Bulatov08:duality,Bulatov07:UAsurvey,Cohen06:handbook,Hell08:survey}.
Classification here is understood with respect to both
computational complexity (i.e.\ membership in a given complexity
class such as P, NL, or L, modulo standard assumptions) and
descriptive complexity (i.e.\ definability of the class of all
positive, or all negative, instances in a given logic).

The best-known classification results in this direction concern the
distinction between polynomial-time solvable and NP-complete CSPs.
For example,  a classical result of Hell and Ne\v set\v ril
(see~\cite{Hell04:book,Hell08:survey}) shows that, for a graph ${\bf
H}$, $\Hom({\bf H})$ (aka ${\bf H}$-colouring)  is tractable if
${\bf H}$ is bipartite or admits a loop, and is NP-complete
otherwise, while Schaefer's dichotomy~\cite{Schaefer78:complexity}
proves that any Boolean CSP is either in P or NP-complete. Recent
work~\cite{Allender09:refining} established a more precise
classification in the Boolean case: if ${\bf T}$ is a structure on
$\{0,1\}$ then $\csp({\bf T})$ is either NP-complete, P-complete,
NL-complete, $\oplus$L-complete, L-complete or in AC$^0$.

Much of the work concerning the descriptive complexity of CSPs is centred around the
database-inspired logic programming language Datalog and its fragments
(see~\cite{Bulatov08:duality,Dalmau05:linear,Egri07:symmetric,Feder98:monotone,Kolaitis08:logical}).
Feder and Vardi initially showed~\cite{Feder98:monotone} that a number of important
tractable cases of $\csp(\bf T)$ correspond to structures for which $\neg \csp(\bf
T)$ (the complement of $\csp(\bf T)$) is definable in Datalog. Similar  ties were
uncovered more recently between the two fragments of Datalog known as linear and
symmetric Datalog and structures ${\bf T}$ for which $\csp({\bf T})$ belongs to NL
and L, respectively~\cite{Dalmau05:linear,Egri07:symmetric}.

Algebra, logic and combinatorics provide three angles of attack
which have fueled progress in this classification
effort~\cite{Bulatov08:duality,Bulatov07:UAsurvey,Cohen06:handbook,Hell04:book,Hell08:survey,Kolaitis08:logical}.
The algebraic approach
(see~\cite{Bulatov07:UAsurvey,Cohen06:handbook}) links the
complexity of $\csp({\bf T})$ to the set of functions that preserve
the relations in ${\bf T}$. In this framework, one associates to
each ${\bf T}$ an algebra $\bbA_{\bf T}$ and exploits the fact that
the properties of $\bbA_{\bf T}$ completely determine the complexity
of $\csp({\bf T})$. This angle of attack was crucial in establishing
key results in the field (see, for
example,~\cite{Barto09:bounded,Bulatov03:conservative,Bulatov07:UAsurvey}).

Tame Congruence Theory, a deep universal-algebraic framework first developed by
Hobby and McKenzie in the mid 80's~\cite{Hobby88:structure}, classifies the local
behaviour of finite algebras into five {\em types} (unary, affine, Boolean, lattice
and semilattice.) It was recently shown
(see~\cite{Bulatov08:duality,Bulatov07:UAsurvey,Larose09:universal}) that there is a
strong connection between the computational and descriptive complexity of $\csp({\bf
T})$ and the set of types that appear in $\bbA_{\bf T}$ and its subalgebras. There
are strong conditions involving types which are sufficient for NL-hardness,
P-hardness and NP-hardness of $\csp({\bf T})$ as well as for inexpressibility of
$\neg \csp({\bf T})$ in Datalog, linear Datalog and symmetric Datalog. These
sufficient conditions are also suspected (and in some cases proved) to be necessary,
under natural complexity-theoretic assumptions. For example, (a) the presence of
unary type is known to imply NP-completeness, while its absence is conjectured to
imply tractability (see~\cite{Bulatov07:UAsurvey}); (b) the absence of unary and
affine types was recently proved to be equivalent to definability in
Datalog~\cite{Barto09:bounded}; (c)~the absence of unary, affine, and semilattice
types is proved necessary, and suspected to be sufficient, for membership in NL and
definability in linear Datalog~\cite{Larose09:universal}; (d)~the absence of all
types but Boolean is proved necessary, and suspected to be sufficient, for
membership in L and definability in symmetric Datalog~\cite{Larose09:universal}. The
strength of evidence varies from case to case and, in particular, the conjectured
algebraic conditions concerning CSPs in NL and L (and, as mentioned above, linear
and symmetric Datalog) still rest on relatively limited
evidence~\cite{Bulatov08:duality,Dalmau05:linear,Dalmau08:symDatalog,Dalmau08:majority,Larose09:universal}.

The aim of the present paper is to show that these algebraic
conditions are indeed sufficient and necessary in the special case
of list $\bH$-colouring for undirected graphs (with possible
loops), and to characterise, in this special case, the dividing
lines in graph-theoretic terms (both via forbidden subgraphs and
through an inductive definition).  One can view the list
$\bH$-colouring problem as a CSP where the template is the
structure $\bH^L$ consisting of the binary (edge) relation of
$\bH$ and all unary relations on $H$ (i.e.\ every subset of $H$).
Tractable list homomorphism problems for general structures were
characterised in~\cite{Bulatov03:conservative} in algebraic terms.
The tractable cases for graphs were described
in~\cite{Feder03:bi-arc} in both combinatorial and (more specific)
algebraic terms; the latter implies, when combined with a recent
result~\cite{Dalmau08:majority}, that in these cases $\neg
\csp({\bH^L})$ definable in linear Datalog and therefore
$\csp(\bH^L)$ is in fact in NL. We complete the picture by
refining this classification and showing that $\csp({\bH^L})$ is
either NP-complete, or NL-complete, or L-complete or in AC$^0$
(and in fact first-order definable). We also remark that the
problem of recognising into which case the problem $\csp({\bH^L})$
falls can be solved in polynomial time.

As we mentioned above, the distinction between NP-complete cases
and those in NL follows from earlier work~\cite{Feder03:bi-arc},
and the situation is similar with distinction between L-hard cases
and those leading to membership in
AC$^0$~\cite{Larose07:FOlong,Larose09:universal}. Therefore, the
main body of technical work in the paper concerns the distinction
between NL-hardness and membership in L. We give two equivalent
characterisations of the class of graphs $\bH$ such that
$\csp({\bH^L})$ is in L. One characterisation is via forbidden
subgraphs (for example, the reflexive graphs in this class are
exactly the $(P_4,C_4)$-free graphs, while the irreflexive ones
are exactly the bipartite $(P_6,C_6)$-free graphs), while the
other is via an inductive definition. The first characterisation
is used to show that graphs outside of this class give rise to
NL-hard problems; we do this by providing constructions witnessing
the presence of a non-Boolean type in the algebras associated with
the graphs. The second characterisation is used to prove positive
results. We first provide operations in the associated algebra
which satisfy certain identities; this allows us to show that the
necessary condition on types is also sufficient in our case. We
also use the inductive definition to demonstrate that the class of
negative instances of the corresponding CSP is definable in
symmetric Datalog, which implies membership of the CSP in L.

\section{Preliminaries}

\subsection{Graphs and relational structures} \label{graph-intro-sec}

In the following we denote the underlying universe of a structure
 $\bS$, $\bT$, ... by its roman equivalent $S$,
$T$,  etc.  A signature is a (finite) set of relation symbols with
associated arities.  Let $\bT$ be a structure of signature $\tau$;
for each relation symbol $R \in \tau$ we denote the corresponding
relation of $\bT$ by $R(\bT)$. Let $\bS$ be a structure of the
same signature. A {\em homo\-morphism} from $\bS$ to $\bT$ is a
map $f$ from $S$ to $T$ such that $f(R(\bS)) \subseteq R(\bT)$ for
each $R \in \tau$. In this case we write $f:\bS\rightarrow\bT$. A
structure $\bT$ is called a {\em core} if every homo\-morphism
from $\bT$ to itself is a permutation on $T$. We denote by
$\CSP(\bT)$ the class of all $\tau$-structures $\bS$ that admit a
homo\-morphism to $\bT$, and by $\neg \CSP(\bT)$ the complement of
this class.

The {\em direct $n$-th power} of a $\tau$-structure $\bT$, denoted $\bT^n$, is
defined to have universe $T^n$ and, for any (say $m$-ary) $R\in \tau$, $({\bf
a}_1\zd {\bf a}_m)\in R(\bT^n)$ if and only if $({\bf a}_1[i]\zd {\bf a}_m[i])\in
R(\bT)$ for each $1\le i\le n$. For a subset $I\subseteq T$, the {\em substructure
induced by $I$ on} $\bT$ is the structure $\bI$ with universe $I$ and such that
$R(\bI)=R(\bT)\cap I^m$ for every $m$-ary $R\in\tau$.

For the purposes of this paper, a {\em graph} is a relational
structure ${\bH} = \langle H;\theta \rangle$ where $\theta$  is a
symmetric binary relation on $H$. The graph $\bH$ is {\em reflexive}
({\em irreflexive}) if $(x,x) \in \theta$ ($(x,x) \not\in \theta$)
for all $x \in H$. Given a graph $\bH$, let $S_1,\dots,S_k$ denote
all subsets of $H$; let $\bH^L$ be the relational structure obtained
from $\bH$ by adding all the $S_i$ as unary relations; more
precisely, let $\tau$ be the signature that consists of one binary
relational symbol $\theta$ and unary symbols $R_i$,  $i =1,\dots,k$.
The $\tau$-structure $\bH^L$ has universe $H$, $\theta(\bH^L)$ is
the edge relation of $\bH$, and $R_i(\bH^L) = S_i$ for all
$i=1,\dots,k$. It is easy to see that $\bH^L$ is a core. We call
$\CSP(\bH^L)$ the {\em list homomorphism problem for $\bH$}. Note
that if $\bG$ is an instance of this problem then $\theta(\bG)$ can
be considered as a digraph, but the directions of the arcs are
unimportant because $\bH$ is undirected. Also, if an element $v\in
G$ is in $R_i(\bG)$ then this is equivalent to $v$ having $S_i$ as
its list, so $\bG$ can be thought of as a digraph with $\bH$-lists.

In \cite{Feder03:bi-arc}, a dichotomy result was proved, identifying bi-arc graphs
as those whose list homomorphism problem is tractable, and others as giving rise to
NP-complete problems. Let $C$ be a circle with two specified points $p$ and $q$. A
bi-arc is a pair of arcs $(N,S)$ such that $N$ contains $p$ but not $q$ and $S$
contains $q$ but not $p$. A  graph $\bH$ is a {\em bi-arc graph} if there is a
family of bi-arcs $\{(N_x,S_x): x \in H\}$ such that, for every $x,y \in H$, the
following hold: (i) if $x$ and $y$ are adjacent, then neither $N_x$ intersects $S_y$
nor $N_y$ intersects $S_x$, and (ii) if $x$ is not adjacent to $y$ then both $N_x$
intersects $S_y$ and  $N_y$ intersects $S_x$.

\subsection{Algebra}

An $n$-ary operation on a set $A$ is a map $f:A^n \rightarrow A$, a {\em projection}
is an operation of the form $e_n^i(x_1,\ldots,x_n)=x_i$ for some $1\le i\le n$.
Given an $h$-ary relation $\theta$ and an
 $n$-ary operation $f$ on the same set $A$, we say that $f$ {\em preserves}
 $\theta$ or that $\theta$ is {\em invariant} under $f$
 if the following holds: given any matrix $M$
of size $h \times n$ whose columns are in $\theta$, applying $f$
to the rows of $M$ will produce an $h$-tuple in $\theta$.

A {\em polymorphism} of a structure $\bT$ is an operation $f$ that
preserves each relation in $\bT$; in this case we also say that
$\bT$ {\em admits} $f$. In other words, an $n$-ary polymorphism of
$\bT$ is simply a homomorphism from $\bT^n$ to $\bT$. With any
structure $\bT$, one associates an algebra $\bbA_{\bT}$ whose
universe is $T$ and whose operations are all polymorphisms of $\bT$.
Given a graph $\bH$, we let $\bbH$ denote the algebra associated
with $\bH^L$. An operation on a set is called {\em conservative} if
it preserves all subsets of the set (as unary relations). So, the
operations of $\bbH$ are the conservative polymorphisms of $\bH$.
Polymorphisms can provide a convenient language when defining
classes of graphs. For example, it was shown in
\cite{Brewster08:nuf} that a graph is a bi-arc graph if and only if
it admits a conservative majority operation where a {\em majority}
operation is a ternary operation $m$ satisfying the identities
$m(x,x,y)=m(x,y,x)=m(y,x,x)=x$.

In order to state some of our results, we will need the notions of a
variety and a term operation. Let $I$ be a signature, i.e. a set of
operation symbols $f$ each of a fixed arity (we use the term
``signature'' for both structures and algebras, this will cause no
confusion). An {\em algebra} of signature $I$ is a pair $\bbA =
\langle A;F\rangle$ where $A$ is a non-empty set (the {\em universe
of $\bbA$}) and $F=\{f^{\bbA}:f \in I\}$ is the set of {\em basic}
operations (for each $f \in I$, $f^{\bbA}$ is an operation on $A$ of
the corresponding arity). The {\em term operations} of $\bbA$ are
the operations built from the operations in $F$ and projections by
using composition. An algebra all of whose (basic or term)
operations are conservative is called a {\em conservative algebra}.
A class of similar algebras (i.e. algebras with the same signature)
 which is closed under
 formation of homomorphic images, subalgebras and direct products is
 called a {\em variety}. The {\em variety generated} by an algebra
 $\bbA$ is denoted by $\cV(\bbA)$, and is
 the smallest variety  containing $\bbA$, i.e. the class of all
 homomorphic images of subalgebras of powers of $\bbA$.

Tame Congruence Theory, as developed in \cite{Hobby88:structure}, is
a powerful tool for the analysis of finite algebras.  Every finite
algebra has a {\em typeset}, which describes (in a certain specified
sense) the local behaviour of the algebra. It contains one or more
of the following 5 {\em types}: (1) the {\em unary} type, (2) the
{\em affine} type, (3) the {\em Boolean} type, (4) the {\em lattice}
type and (5) the {\em semilattice} type. The numbering of the types
is fixed, and they are often referred to by their numbers. Simple
algebras, i.e. algebras without non-trivial proper homomorphic
images, admit a unique type; the prototypical examples are: any
2-element algebra whose basic operations are all unary has type 1. A
finite vector space has type 2. The 2-element Boolean algebra has
type 3. The 2-element lattice is the 2-element algebra with two
binary operations $\langle \{0,1\};\vee,\wedge \rangle$: it has type
4. The 2-element semilattices are the 2-element algebras with a
single binary operation $\langle \{0,1\};\wedge \rangle$ and
$\langle \{0,1\};\vee \rangle$: they have type 5.     The typeset of
a variety $\cV$, denoted $typ(\cV)$, is simply the union of typesets
of the algebras in it. We will be mostly interested in type-omitting
conditions for varieties of the form $\cV(\bbA_{\bT})$, and
Corollary~3.2 of~\cite{Valeriote09:intersection} says that in this
case it is enough to consider the typesets of $\bbA_{\bT}$ and its
subalgebras.

On the intuitive level, if $\bT$ is a core structure then the typeset
$typ(\cV(\bbA_{\bT}))$ contains crucial information about the kind of relations that
$\bT$ can or cannot simulate, thus implying lower/upper bounds on the complexity of
$\CSP(\bT)$.
 For our purposes here, it will not be necessary to delve further
 into
the technical aspects of types and typesets. We only note that there
is a very tight connection between the kind of equations that are
satisfied by the algebras in a variety and the types that are {\em
admitted} or {\em omitted} by a variety, i.e. those  types that do
or do not appear in the typesets of algebras in the
variety~\cite{Hobby88:structure}.

In this paper, we use ternary operations $f_1,\dots,f_n$
satisfying the following identities:
\begin{eqnarray}
x & = & f_1(x,y,y) \label{n-perm-ops1} \\  f_i(x,x,y) & = & f_{i+1}(x,y,y) \mbox{ for all $i=1,\ldots n-1$} \label{n-perm-ops2}\\
 f_n(x,x,y) & = & y. \label{n-perm-ops3}
\end{eqnarray}

The following lemma contains some type-omitting results that we use in this paper.

\begin{lemma} \cite{Hobby88:structure} \label{algebra_lemma}
A finite algebra $\bbA$ has term operations $f_1,\ldots,f_n$, for
some $n\ge 1$, satisfying
identities~(\ref{n-perm-ops1})--(\ref{n-perm-ops3})
 if and only if the variety
$\cV(\bbA)$ omits types 1, 4 and 5.

\noindent If a finite algebra $\bbA$ has a majority term operation then $\cV(\bbA)$
omits types 1, 2 and 5.
\end{lemma}

We remark in passing that operations satisfying
identities~(\ref{n-perm-ops1})--(\ref{n-perm-ops3}) are also known
to characterise a certain algebraic (congruence) condition called
$(n+1)$-permutability~\cite{Hobby88:structure}.


\subsection{Datalog}

Datalog is a query and rule language for deductive databases
(see~\cite{Kolaitis08:logical}). A Datalog program \datprog\ over
a (relational) signature $\tau$ is a finite set of rules of the
form $h \leftarrow b_1 \wedge \ldots \wedge b_m$ where $h$ and
each $b_i$ are atomic formulas $R_j(v_1,...,v_k)$. We say that $h$
is the {\em head} of the rule and that $b_1 \wedge \ldots \wedge
b_m$ is its {\em body}. Relational predicates $R_j$ which appear
in the head of some rule of $\cD$ are called {\em intensional
database predicates} ({\em IDB}s) and are not part of the
signature $\tau$. All other relational predicates are called {\em
extensional database predicates} ({\em EDB}s) and are in $\tau$.
So, a Datalog program is a recursive specification of IDBs (from
EDBs).

A rule of \datprog\ is {\em linear} if its body contains at most one IDB and is {\em
non-recursive} if its body contains only EDBs. A linear but recursive rule is of the
form $I_1(\bar{x}) \leftarrow I_2(\bar{y}) \wedge  E_1(\bar{z}_1) \wedge  \ldots
\wedge E_k(\bar{z}_k)$ where $I_1, I_2$ are IDBs and the $E_i$ are EDBs (note that
the variables occurring in $\bar{x}, \bar{y}, \bar{z}_i$ are not necessarily
distinct). Each such rule has a {\em symmetric} $I_2(\bar{y}) \leftarrow
I_1(\bar{x}) \wedge E_1(\bar{z}_1) \wedge \ldots  \wedge E_k(\bar{z}_k).$ A Datalog
program is {\em non-recursive} if all its rules are non-recursive, {\em linear} if
all its rules are linear and {\em symmetric} if it is linear and if the symmetric of
each recursive rule of \datprog\ is also a rule of \datprog.

A Datalog program \datprog\ takes a $\tau$-structure $\bf A$ as
input and returns a structure \datprog$(\bf A)$ over the signature
$\tau' = \tau \cup \{I:I$ is an IDB in \datprog$\}$. The relations
corresponding to $\tau$ are the same as in $\bf A$, while the new
relations are recursively computed by \datprog\ , with semantics
naturally obtained via least fixed-point of monotone operators. We
also want to view a Datalog program as being able to accept or
reject an input $\tau$-structure and this is achieved by choosing
one of the IDBs of \datprog\ as the {\em goal predicate}: the
$\tau$-structure $\bf A$ is {\em accepted by} \datprog\ if the goal
predicate is non-empty in $\datprog(\bf A)$. Thus every Datalog
program with a goal predicate defines a class of structures - those
that are accepted by the program.

When using Datalog to study $\CSP(\bT)$, one usually speaks of the definability of
$\neg \CSP(\bT)$ in Datalog (i.e. by a Datalog program) or its fragments (because
any class definable in Datalog must be closed under extension). Examples of CSPs
definable in Datalog and its fragments can be found, e.g.,
in~\cite{Bulatov08:duality,Egri07:symmetric}. As we mentioned before, any problem
$\CSP(\bT)$ is tractable if its complement is definable in Datalog, and all such
structures were recently identified in~\cite{Barto09:bounded}. Definability of $\neg
\CSP(\bT)$ in linear (symmetric) Datalog implies that $\CSP(\bT)$ belongs to NL and
L, respectively~\cite{Dalmau05:linear,Egri07:symmetric}. As we discussed in
Section~\ref{intro}, there is a connection between definability of CSPs in Datalog
(and its fragments) and the presence/absence of types in the corresponding algebra
(or variety).

Note that it follows from Lemma~\ref{algebra_lemma} and from the
results in~\cite{Larose09:universal,Larose06:bounded} that if, for a
core structure $\bT$, $\neg \CSP(\bT)$ is definable in symmetric
Datalog then $\bT$ must admit, for some $n$, operations satisfying
identities~(\ref{n-perm-ops1})--(\ref{n-perm-ops3}). Moreover, with
the result of~\cite{Barto09:bounded}, a conjecture
from~\cite{Larose09:universal} can be restated as follows: for a
core structure $\bT$, if $\neg \CSP(\bT)$ is definable in Datalog
and, for some $n$, $\bT$ admits operations
satisfying~(\ref{n-perm-ops1})--(\ref{n-perm-ops3}), then $\neg
\CSP(\bT)$ is definable in symmetric Datalog. This conjecture is
proved in~\cite{Dalmau08:symDatalog} for $n=1$.

\section{A class of graphs} \label{graph-sec}

In this section, we give combinatorial characterisations of a
class of graphs whose list homomorphism problem will turn out to
belong to L.

Let $\bH_1$ and $\bH_2$ be bipartite irreflexive graphs, with
colour classes $B_1$, $T_1$ and $B_2$ and $T_2$ respectively, with
$T_1$ and $B_2$ non-empty. We define the {\em special sum} $\bH_1
\odot \bH_2$ (which depends on the choice of the $B_i$ and $T_i$)
as follows: it is the graph obtained from the disjoint union of
$\bH_1$ and $\bH_2$ by adding all possible edges between the
vertices in $T_1$ and $B_2$. Notice that we can often decompose a
bipartite graph in several ways, and even choose $B_1$ or $T_2$ to
be empty. We say that an irreflexive graph $\bH$ is {\em a special
sum} or {\em expressed as a special sum} if there exist two
bipartite graphs and a choice of colour classes on each such that
$\bH$ is isomorphic to the special sum of these two graphs.

\begin{defi} Let $\mathcal K$ denote the smallest class of irreflexive graphs
containing the one-element graph and closed under (i) special sum
and (ii) disjoint union. We call the graphs in $\mathcal K$ {\em
basic irreflexive}.
\end{defi}

The following result gives a characterisation of basic irreflexive
graphs in terms of forbidden subgraphs:

\begin{lemma}\label{characterize_good_irreflexive} Let $\bH$ be an irreflexive graph. Then the following
conditions are equivalent:
\begin{enumerate}
\item $\bH$ is basic irreflexive; \item $\bH$ is bipartite,
contains no induced 6-cycle, nor any induced path of
length~5.\end{enumerate}
 \end{lemma}

We shall now describe our main family of graphs, first by
forbidden induced subgraphs, and then in an inductive manner.

\begin{defi} \label{forb-def} Define the class $\cL$ of graphs as follows: a graph
$\bH$ belongs to $\cL$ if it contains none of the following as an induced subgraph:
\begin{enumerate}

\item the reflexive path of length 3 and the reflexive 4-cycle;
\item the irreflexive cycles of length 3, 5 and 6, and the
irreflexive path of length 5; \item ${\bf B1}$, ${\bf B2}$, ${\bf
B3}$, ${\bf B4}$, ${\bf B5}$ and ${\bf B6}$ (see Figure
\ref{badguys}.)
\end{enumerate}
\end{defi}
\begin{center}

\begin{figure}
\scalebox{1}{
\psset{yunit=0.7}
\begin{pspicture}(1,1)(10,6)
\uput[u](1,0.8){$\mathbf{B1}$} \uput[u](2,0.8){$\mathbf{B2}$}
\uput[u](3,0.8){$\mathbf{B3}$} \uput[u](4,0.8){$\mathbf{B4}$}
\uput[u](6,0.8){$\mathbf{B5}$} \uput[u](8.5,0.8){$\mathbf{B6}$}

\uput[r](1,4){$c$} \uput[r](1,3){$b$} \uput[r](1,2){$a$}
\uput[r](2,4){$c$} \uput[r](2,3){$b$} \uput[r](2,2){$a$}
\uput[r](3,5){$d$} \uput[r](3,4){$c$} \uput[r](3,3){$b$}
\uput[r](3,2){$a$} \uput[r](4,6){$e$} \uput[r](4,5){$d$}
\uput[ur](4,4){$c$} \uput[r](4,3){$b$} \uput[dr](4,2){$a$}

\uput[l](5.5,4){$a'$} \uput[l](5.5,3){$b'$} \uput[l](5.5,2){$c'$}
\uput{10pt}[20](6.5,4){$a$} \uput[ul](6.5,3){$b$}
\uput{10pt}[-20](6.5,2){$c$}

\uput[l](8,4){$a'$} \uput[l](8,3){$b'$} \uput[l](8,2){$c'$}
\uput{10pt}[20](9,4){$a$} \uput[ul](9,3){$b$}
\uput{10pt}[-20](9,2){$c$}

\cnode*(1,4){2.0pt}{0} \cnode*(1,3){2.0pt}{1}
\cnode*(1,2){2.0pt}{2} \ncline{0}{1} \ncline{1}{2}
\nccircle[angleA=90]{0}{0.15}

\cnode*(2,4){2.0pt}{0} \cnode*(2,3){2.0pt}{1}
\cnode*(2,2){2.0pt}{2} \ncline{0}{1} \ncline{1}{2}
\nccircle[angleA=90]{0}{0.15} \nccircle[angleA=90]{2}{0.15}

\cnode*(3,5){2.0pt}{0} \cnode*(3,4){2.0pt}{1}
\cnode*(3,3){2.0pt}{2} \cnode*(3,2){2.0pt}{3} \ncline{0}{1}
\ncline{1}{2} \ncline{2}{3} \nccircle[angleA=90]{0}{0.15}
\nccircle[angleA=90]{1}{0.15} \nccircle[angleA=90]{2}{0.15}

\cnode*(4,6){2.0pt}{0} \cnode*(4,5){2.0pt}{1}
\cnode*(4,4){2.0pt}{2} \cnode*(4,3){2.0pt}{3}
\cnode*(4,2){2.0pt}{4} \ncline{0}{1} \ncline{1}{2} \ncline{2}{3}
\ncline{3}{4} \nccircle[angleA=90]{1}{0.15}
\nccircle[angleA=90]{2}{0.15}
 \nccurve{2}{4}

\cnode*(6.5,4){2.0pt}{0} \cnode*(6.5,3){2.0pt}{1}
\cnode*(6.5,2){2.0pt}{2} \ncline{0}{1} \ncline{1}{2}
\nccircle{0}{0.15}\nccircle[angleA=-90]{1}{0.15}
\nccircle[angleA=180]{2}{0.15} \cnode*(5.5,4){2.0pt}{3}
\cnode*(5.5,3){2.0pt}{4} \cnode*(5.5,2){2.0pt}{5} \ncline{2}{5}
\ncline{1}{5} \ncline{1}{4} \ncline{0}{4}\ncline{0}{3}
\nccurve{2}{0}

\cnode*(9,4){2.0pt}{0} \cnode*(9,3){2.0pt}{1}
\cnode*(9,2){2.0pt}{2} \ncline{0}{1} \ncline{1}{2}
\nccircle{0}{0.15}\nccircle[angleA=-90]{1}{0.15}
\nccircle[angleA=180]{2}{0.15} \cnode*(8,4){2.0pt}{3}
\cnode*(8,3){2.0pt}{4} \cnode*(8,2){2.0pt}{5} \ncline{2}{5}
\ncline{1}{5} \ncline{1}{4} \ncline{0}{4}\ncline{0}{3}
\ncline{2}{3} \nccurve{2}{0}
\end{pspicture}}

\caption{The forbidden mixed graphs.} \label{badguys}
\end{figure}
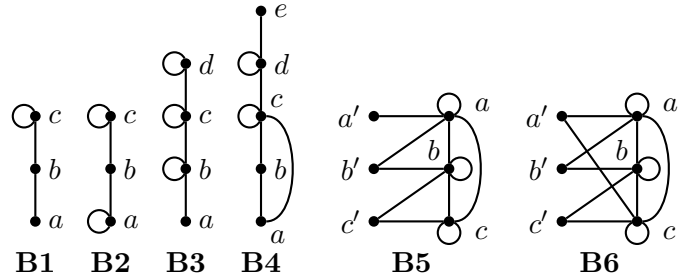
\end{center}

\vspace{-0.5cm}

 We will now characterise the class $\cL$ in an
inductive manner.

\begin{defi} A connected graph $\bH$ is {\em basic} if  either (i) $\bH$ is a single loop,  or (ii) $\bH$ is
a  basic irreflexive graph, or (iii) $\bH$ is obtained from a
basic irreflexive graph $\bH_1$ with colour classes $B$ and $T$ by
adding every edge (including loops) of the form $\{t,t'\}$ where
$t,t' \in T$.
\end{defi}

\begin{defi}
Given two vertex-disjoint graphs $\bH_1$ and $\bH_2$, the  {\em adjunction of
$\bH_1$ to $\bH_2$}  is the graph $\bH_1 \oslash \bH_2$ obtained by taking the
disjoint union of the two graphs, and adding every edge of the form $\{x,y\}$ where
$x$ is a loop in $\bH_1$ and $y$ is a vertex of $\bH_2$.
\end{defi}

\begin{lemma} \label{reflexive_lemma} Let $\cL_R$ denote the class of reflexive graphs in $\cL$.
 Then $\cL_R$ is the smallest class
 $\cD$ of reflexive graphs such that:
\begin{enumerate}
\item $\cD$ contains the one-element graph; \item $\cD$ is closed
under disjoint union; \item if $\bH_1$ is a single loop and $\bH_2
\in \cD$ then $\bH_1 \oslash \bH_2 \in \cD$.
\end{enumerate}   \end{lemma}

Lemma~\ref{reflexive_lemma}  states that the reflexive graphs
avoiding the path of length 3 and the 4-cycle are precisely those
 constructed from the one-element loop using disjoint union and
 adjunction of a universal vertex. These graphs can also be described by the following property: every
connected induced subgraph  of size at
 most 4 has a universal vertex. These graphs have been
 studied previously as those with NLCT width 1, which were
 proved to be exactly the
 trivially perfect graphs~\cite{Gurski06:co-graphs}. Our result provides an alternative
proof of the equivalence of these conditions.

\begin{theorem}\label{characterize_good_graphs} The class $\cL$ is
the smallest class $\ccC$ of graphs such that:
\begin{enumerate}
\item $\ccC$ contains the basic graphs; \item $\ccC$ is closed under
disjoint union; \item if $\bH_1$ is a basic graph and $\bH_2 \in
\ccC$ then $\bH_1 \oslash \bH_2 \in \ccC$.
\end{enumerate}
 \end{theorem}

 \proof
 We start by showing that every basic graph is in $\cL$, i.e. that
 a basic graph does not contain any of the forbidden
 graphs. If $\bH$ is a single loop or a basic irreflexive graph, then this is immediate.
 Otherwise $\bH$ is obtained from a basic irreflexive graph
$\bH_1$ with colour classes $B$ and $T$ by adding every edge of the form $(t_1,t_2)$
where $t_i \in T$. In particular,  the loops form a clique and no edge connects two
non-loops; it is clear in that case that $\bH$ contains none of ${\bf B1}$, ${\bf
B2}$, ${\bf B3}$, ${\bf B4}$. On the other hand if  $\bH$ contains ${\bf B5}$ or
${\bf B6}$, then $\bH_1$ contains the path of length 5 or the 6-cycle, contradicting
the fact that $\bH_1$ is basic.

Next we show that $\cL$ is closed under disjoint union and adjunction of basic
graphs. It is obvious that the disjoint union of graphs that avoid the forbidden
graphs will also avoid these. So suppose that an adjunction $\bH_1 \oslash \bH_2$,
where $\bH_1$ is a basic graph, contains an induced forbidden graph $\bB$ whose
vertices are neither all in $H_1$ nor $H_2$;  without loss of generality $H_1$
contains at least one loop,  its loops form a clique and  none of its edges connects
two non-loops. It is then easy to verify that $\bB$ contains both loops and
non-loops. Because the other cases are similar, we  prove only that $\bB$ is not
${\bf B3}$: since vertex $d$ is not adjacent to $a$ it must be in $\bH_2$, and
similarly for $c$. Since $b$ is not adjacent to $d$ it must also be in $\bH_2$;
since non-loops of $\bH_1$ are not adjacent to elements of $\bH_2$ it follows that
$a$ is in $\bH_2$ also, a contradiction.

Now we must show that every graph in $\cL$ can be obtained from
the basic graphs by disjoint union and adjunction of basic graphs.
Suppose this is not the case. If $\bH$ is a counterexample of
minimum size, then obviously it is connected, and it contains at
least one loop for otherwise it is a basic irreflexive graph. By
Lemma \ref{reflexive_lemma}, $\bH$ also contains at least one
non-loop.

For $a \in H$ let $N(a)$ denote its set of neighbours. Let
$\bR(\bH)$ denote the subgraph of $\bH$ induced by its set $R(H)$ of
loops, and let $\bJ(\bH)$ denote the subgraph induced by $J(H)$, the
set of non-loops of $\bH$. Since $\bH$ is connected and neither
${\bf B1}$ nor ${\bf B2}$  is an induced subgraph of $\bH$, the
graph $\bR(\bH)$ is also connected, and furthermore every vertex in
$J(H)$ is adjacent to some vertex in $R(H)$. By
Lemma~\ref{reflexive_lemma}, we know that $\bR(\bH)$ contains at
least one universal vertex: let $U$ denote the (non-empty) set of
universal vertices of $\bR(\bH)$. Let $J$ denote the set of all $a
\in J(H)$ such that $N(a) \cap R(H) \subseteq U$. Let us show that
$J\ne \emptyset$. For every $u\in U$, there is $w\in J(H)$ not
adjacent to $u$ because otherwise $\bH$ is obtained by adjoining $u$
to the rest of $\bH$, a contradiction with the choice of $\bH$. If
this $w$ has a neighbour $r\in R(H)\setminus U$ then there is some
$s\in R(H)\setminus U$ not adjacent to $r$, and the graph induced by
$\{w,u,s,r\}$ contains ${\bf B2}$ or ${\bf B3}$, a contradiction.
Hence, $w\in J$. Let $\bS$ denote the subgraph of $\bH$ induced by
$U \cup J$. The graph $\bS$ is connected. We claim that the
following properties also hold:

\begin{enumerate}
\item if $a$ and $b$ are adjacent non-loops, then $N(a) \cap U =
N(b) \cap U$; \item if $a$ is in a connected component of the subgraph of $\bS$
induced by $J$ with more than one vertex, then for any other $b \in J$, one of $N(a)
\cap U, N(b) \cap U$ contains the other.
\end{enumerate}

The first statement holds because ${\bf B1}$ is forbidden, and the
second  follows from the first because ${\bf B4}$ is also forbidden.
Let $J_1,\dots,J_k$ denote the different connected components of $J$
in $\bS$. By (1) we may let $N(J_i)$ denote the set of common
neighbours of members of $J_i$ in $U$. By (2), we can re-order the
$J_i$'s so that for some $1\le m\le k$ we have $N(J_i) \subseteq
N(J_j)$ for all $i \leq m$ and all $j > m$, and, in addition, we
have $m=1$ or $|J_i|=1$ for all $1\le i\le m$. Let $\bB$ denote the
subgraph of $\bS$ induced by $B = \bigcup_{i=1}^m{(J_i \cup
N(J_i))}$, and let $\bC$ be the subgraph of $\bH$ induced by $H
\setminus B$. We claim that $\bH = \bB \oslash \bC$. For this, it
suffices to show that every element in $\bigcup_{i=1}^m N(J_i) $ is
adjacent to every non-loop $c \in C$. By construction this holds if
$c \in J \cap C$. Now suppose this does not hold: then some $x \in
J(H) \setminus J$ is not adjacent to some $y \in N(J_i)$ for some $i
\leq m$. Since $x \not\in J$ we may find some $z \in R(H) \setminus
U$ adjacent to $x$; it is of course also adjacent to $y$. Since $z
\not\in U$ there exists some $z' \in R(H) \setminus U$ that is not
adjacent to $z$, but it is of course adjacent to $y$.  If $x$ is
adjacent to $z'$, then $\{x,z,z'\}$ induces a subgraph isomorphic to
${\bf B2}$, a contradiction. Otherwise, $\{x,z,y,z'\}$ induces a
subgraph isomorphic to ${\bf B3}$, also a contradiction.

If every $J_i$ with $i \leq m$ contains a single
element, notice that $\bB$ is a basic graph: indeed, removing all
edges between its loops yields a bipartite irreflexive graph which
contains neither the path of length 5 nor the 6-cycle, since $\bB$
contains neither ${\bf B5}$ nor ${\bf B6}$. Since this contradicts
our hypothesis on $\bH$, we conclude that $m=1$.   But this means
that $N(J_1)$ is a set of universal vertices in $\bH$. Let $u$ be
such a vertex and let $D$ denote its complement in $\bH$: clearly
$\bH$ is obtained as the adjunction of the single loop $u$ to $D$,
contradicting our hypothesis. This concludes the proof. \qed

\section{Classification results}

Recall the standard numbering of types: (1)  {\em unary}, (2) {\em
affine} , (3)  {\em Boolean}, (4) {\em lattice}  and (5)  {\em
semilattice}. We will need the following auxiliary result (which
is well known). Note that the assumptions of this lemma
effectively say that $\csp(\bT)$ can simulate the graph
$k$-colouring problem (with $k=|U|$) or the directed
$st$-connectivity problem.

\begin{lemma} \label{disequality-order-lemma}
Let $\bS, \bT$ be structures, let $s_1,s_2 \in S$, and let
$R=\{(f(s_1),f(s_2))\mid f: \bS \rightarrow \bT\}$.

\begin{enumerate}
\item If $R=\{(x,y)\in U^2 \mid x\ne y\}$ for some subset $U
\subseteq T$ with $|U|\ge 3$ then $\cV(\bbA_{\bT})$ admits type 1.

\item If $R=\{(t,t),(t,t'),(t',t')\}$ for some distinct $t,t'\in
T$ then $\cV(\bbA_{\bT})$ admits at least one of the types 1, 4, 5.
\end{enumerate}
\end{lemma}

\begin{proofsketch} The assumption of this lemma implies that $\bbA_{\bT}$
has a subalgebra (induced by $U$ and $\{t,t'\}$, respectively) such
that all operations of the subalgebra preserve the relation $R$. It
is well-known (see, e.g.,~\cite{Hell04:book}) that all operations
preserving the disequality relation on $U$ are essentially unary,
while it is easy to check that the order relation on a 2-element set
cannot admit operations satisfying
identities~(\ref{n-perm-ops1})--(\ref{n-perm-ops3}), so one can use
Lemma~\ref{algebra_lemma}.
\end{proofsketch}

The following lemma connects the characterisation of bi-arc graphs
given in~\cite{Brewster08:nuf} with a type-omitting condition.

\begin{lemma} \label{taylor_ops}
Let $\bH$ be a graph. Then the following conditions are equivalent:
\begin{enumerate}
    \item the variety ${\mathcal V}(\bbH)$ omits type 1;
    \item the graph $\bH$ admits a conservative majority operation;
    \item the graph $\bH$ is a bi-arc graph.
\end{enumerate}    \end{lemma}

The results summarised in the following theorem are known (or easily follow from
known results, with a little help from Lemma~\ref{taylor_ops}).

\begin{theorem} \label{oldstuff}
Let $\bH$ be a graph.
\begin{itemize}
\item If $typ({\mathcal V}(\bbH))$ admits type 1,  then $\neg
\CSP(\bH^L)$ is not expressible in Datalog and $\CSP(\bH^L)$ is
$\mathrm{NP}$-complete (under first-order reductions); \item if
$typ({\mathcal V}(\bbH))$ omits type 1 but admits type 4 then
$\neg \CSP(\bH^L)$ is not expressible in symmetric Datalog but is
expressible in linear Datalog, and $\CSP(\bH^L)$ is
$\mathrm{NL}$-complete (under first-order reductions.)
\end{itemize}
\end{theorem}

\proof The first statement is shown in~\cite{Larose09:universal}. If the variety
omits type 1, then $\bH^L$ admits a majority operation by Lemma \ref{taylor_ops} and
then $\neg \CSP(\bH^L)$ is expressible in linear Datalog
by~\cite{Dalmau08:majority}; in particular the problem is in NL. If, furthermore,
the variety admits type 4, then $\neg \CSP(\bH^L)$ is not expressible in symmetric
Datalog and is NL-hard by results in \cite{Larose09:universal}. \qed

By Lemma~\ref{algebra_lemma}, the presence of a majority operation in $\bbH$ implies
that $typ({\mathcal V}(\bbH))$ can contain only types 3 and 4. Type 4 is dealt with
in Theorem~\ref{oldstuff},
 so it remains to investigate graphs
{\bf H} with $typ({\mathcal V}(\bbH)) = \{3\}$.

The next theorem is the main result of this paper.

\begin{theorem}\label{main}  Let $\bH$ be a graph. Then the following
conditions are equivalent:

\begin{enumerate}

    \item $\bH$ admits conservative operations
    satisfying~(\ref{n-perm-ops1})--(\ref{n-perm-ops3}) for $n=3$;

    \item $\bH$ admits conservative operations
    satisfying~(\ref{n-perm-ops1})--(\ref{n-perm-ops3}) for some $n\ge 1$;
    \item $typ({\mathcal V}(\bbH)) = \{3\}$;
    \item $\bH \in \cL$;
    \item $\neg \CSP(\bH^L)$ is definable in symmetric Datalog.
\end{enumerate}
If the above holds then $\CSP(\bH^L)$  is in the complexity class $\mathrm{L}$.
\end{theorem}

\begin{proofsketch} (1) $\Rightarrow$ (2) is trivial. If (2) holds then
by Lemma \ref{algebra_lemma} ${\mathcal V}(\bbH)$ omits types 1,
4, and 5. By Lemma \ref{taylor_ops}, $\bH$ admits a majority
operation, so Lemma \ref{algebra_lemma} implies that ${\mathcal
V}(\bbH)$ also omits type 2; hence (3) holds. Implication
(3)$\Rightarrow$(4) is the content of Lemma~\ref{3to4-lem} below,
and (5) implies (3) by a result of \cite{Larose09:universal}. By
using Theorem~\ref{characterize_good_graphs}, one can show that
(4) implies both (1) and (5). Finally, definability in symmetric
Datalog implies membership in L by~\cite{Egri07:symmetric}.
\end{proofsketch}

\begin{lemma}\label{3to4-lem}
If $\bH\not\in\cL$ then $typ({\mathcal V}(\bbH)) \ne \{3\}$.
\end{lemma}

\proof By Theorem~9.15 of~\cite{Hobby88:structure}, $typ({\mathcal
V}(\bbH)) = \{3\}$ if and only if $\bH$ admits a sequence of
conservative operations satisfying certain identities (in the spirit
of~(\ref{n-perm-ops1})--(\ref{n-perm-ops3})). By conservativity,
such operations can be restricted to any subset of $H$ while
satisfying the same identities, so the property $typ({\mathcal
V}(\bbH)) = \{3\}$ is inherited by induced subgraphs. It follows
that it is enough to prove this lemma for the forbidden graphs from
Definition~\ref{forb-def}.

For the irreflexive odd cycles, the lemma follows immediately from the main results
of~\cite{Barto09:digraphs,Maroti08:existence}. The proof of Theorem~3.1
of~\cite{Feder99:list} shows that the conditions of
Lemma~\ref{disequality-order-lemma}(1) are satisfied by (some $\bS,s_1,s_2$ and)
$\bT=\bF^L$ where $\bF$ is the irreflexive 6-cycle. One can check that the
reflexive 4-cycle is not a bi-arc graph, so we can apply Lemma~\ref{taylor_ops} in
this case.

For the remaining forbidden graphs $\bF$ from Definition~\ref{forb-def}, we use
Lemma~\ref{disequality-order-lemma}(2) with $\bT=\bF^L$. In each case, the binary
relation of the structure $\bS$ will be a short undirected path, and $s_1,s_2$ will
be the endpoints of the path. We will represent such a structure $\bS$ by a sequence
of subsets of $F$ (indicating lists assigned to vertices of the path). It can be
easily checked that, in each case, the relation $R$ defined as in
Lemma~\ref{disequality-order-lemma} is of the required form.

If $\bF$ is the reflexive path of length 3, say $a-b-c-d$, then $\bS=ac-bc-ad-ac$.
If $\bF$ is the irreflexive path of length 5, say $a-b-c-d-e-f$ then
$\bS=ae-bd-ce-bf-ae$. For graphs ${\bf B1}-{\bf B6}$, we use notation from
Fig.~\ref{badguys}. For ${\bf B1}$, $\bS=bc-bc-ab-ab-bc$. For  ${\bf B2}$,
$\bS=bc-ac-ab-bc$. For ${\bf B3}$, $\bS=bc-ad-bd-bc$. For ${\bf B4}$,
$\bS=ae-bd-cd-ae$. Finally, for both ${\bf B5}$ and ${\bf B6}$,
$\bS=ac-b'c'-ab-a'c'-ac$.
 \qed

For completeness' sake, we describe graphs whose list homomorphism
problem is definable in first-order logic (equivalently, is in
AC$^0$, see \cite{Bulatov08:duality}.) By results in
\cite{Larose09:universal}, any problem $\CSP(\bT)$ is either
first-order definable or L-hard under FO reductions. Hence, it
follows from Theorem~\ref{main} that, for a graph $\bH \in \cL$, the
list homomorphism problem for $\bH$ is either first-order definable
or L-complete.

We need the following characterisation of structures whose CSP is
first-order definable \cite{Larose07:FOlong}. Let $\bT$ be a
relational structure and let $a,b \in T$. We say that $b$ {\em
dominates} $a$ in $\bT$ if for any relation $R$ of $\bT$, and any
tuple $\overline{t} \in R$, replacement of any occurrence of $a$ by
$b$ in $\overline{t}$ will yield a tuple of $R$. Recall the
definition of a direct power of a structure from
Subsection~\ref{graph-intro-sec}. If $\bT$ is a relational
structure, we say that the structure $\bT^2$ {\em dismantles to the
diagonal} if there exists a sequence of elements $\{a_0,\dots,a_n\}
= T^2 \setminus \{(a,a):a \in T\}$ such that, for all $0 \leq i \leq
n$, $a_i$ is dominated in $\bT_i$, where $\bT_0= \bT^2$ and $\bT_i$
is the substructure of $\bT^2$ induced by $T^2 \setminus
\{a_0,\dots,a_{i-1}\}$ for $i > 0$.

\begin{lemma}[\cite{Larose07:FOlong}]\label{dismantle}
Let $\bT$ be a core relational structure. Then $\CSP(\bT)$ is
first-order definable if and only if $\bT^2$ dismantles to the
diagonal.
\end{lemma}

\begin{theorem} \label{FO_symmetric} Let $\bH$ be a graph. Then $\CSP(\bH^L)$ is
first-order definable if and only if $\bH$ has the following form: $H$ is the
disjoint union of two sets $L$ and $N$ such that (i) $L$ is the set of loops of
$\bH$ and induces a complete graph, (ii) $N$ is the set of non-loops of $\bH$ and
induces a graph with no edges, and (iii) $N=\{x_1,\dots,x_m\}$ can be ordered so
that the neighbourhood of $x_i$ is contained in the neighbourhood of $x_{i+1}$ for
all $1 \leq i \leq m-1$.
\end{theorem}

\proof We first prove that conditions (i) and (ii) are necessary. Notice that if
$\csp(\bH^L)$ is first-order definable then so is $\csp(\bK^L)$  for any induced
substructure $\bK$ of $\bH$. Let $x$ and $y$ be distinct vertices of $\bH$ and let
$\bK^L$ be the substructure of $\bH^L$ induced by $\{x,y\}$. If $x$ and $y$ are
non-adjacent loops, then $\theta(\bK) = \{(x,x),(y,y)\}$ the equality relation on
$\{x,y\}$; if $x$ and $y$ are adjacent non-loops, then $\theta(\bK) =
\{(x,y),(y,x)\}$, the adjacency relation of the complete graph on 2 vertices. It is
well known (and can be easily derived from Lemma~\ref{dismantle}) that neither of
these classes $\csp(\bK^L)$ is first-order definable. It follows that the loops of
$\bH$ induce a complete graph and the non-loops induce a graph with no edges.

Now we prove (iii) is necessary. Suppose for a contradiction that
there exist distinct elements $x$ and $y$ of $N$ and elements $n$
and $m$ of $L$ such that $m$ is adjacent to $x$ but not to $y$,
and $n$ is adjacent to $y$ but not to $x$. Then $\CSP(\bG)$ is
first-order definable, where $\bG$ is the substructure
 of $\bH^L$ induced by $\{x,y,m,n\}$. By Lemma \ref{dismantle}, $\bG^2$ dismantles to the
diagonal. Then $(x,y)$ must be dominated by one of $(x,x)$,
$(y,x)$ or $(y,y)$, since domination respects the unary relation
$\{x,y\}^2$ (on $G^2$). But $(m,n)$ is a neighbour of $(x,y)$ and
none of the other three, a contradiction.

For the converse: we show that we can dismantle $(\bH^L)^2$ to the
diagonal. Let $x \in H$: then $(x_1,x)$ and $(x,x_1)$ are
dominated by $(x,x)$. Suppose that we have dismantled every
element containing a coordinate equal to $x_i$ with $i \leq j-1$:
if $x$ is any element of $H$ such that the elements $(x_j,x)$ and
$(x,x_j)$ remain, then either $x$ is a loop or $x=x_k$ with $k
\geq j$; in any case the elements $(x_j,x_k)$ and $(x_k,x_j)$ are
dominated by $(x,x)$. In this way we can remove all pairs $(x,y)$
with one of $x$ or $y$ a non-loop. For the remaining pairs, notice
that if $u$ and $v$ are any loops then  $(u,v)$ is dominated (in
what remains of $(\bH^L)^2$) by $(u,u)$. \qed

Finally, given a graph $\bH$, it can be decided in
polynomial time which of the different cases delineated in
Theorems~\ref{oldstuff},~\ref{main},~\ref{FO_symmetric} the list
homomorphism problem for $\bH$ satisfies. Indeed, it is known  that bi-arc graphs can be recognised in polynomial time (see \cite{Feder03:bi-arc}).\nopagebreak[4] Assume that $\bH$ is a bi-arc graph:
the forbidden substructure definition of the class $\cL$ gives an
$AC^0$ algorithm to recognise them; and those graphs whose list
homomorphism problem is first-order definable can be recognised in
polynomial time by results of~\cite{Larose07:FOlong}.


\end{document}